\documentclass[11pt,a4paper]{elsarticle}
\pdfoutput=1
\usepackage{amsmath,amssymb,enumerate,graphicx,caption,subcaption,fancyhdr}
\usepackage[usenames,dvipsnames,svgnames,table]{xcolor}
\usepackage[T1]{fontenc}

\newcommand{\ud}{\mathrm{d}}

\title{A Remark on Polar Noncommutativity}

\author{Andrew Iskauskas}
\ead{andrew.iskauskas@durham.ac.uk}
\address{Department of Mathematical Sciences, Durham University \\
Lower Mountjoy, Stockton Road, Durham DH1 3LE, UK}

\begin{document}

\begin{abstract}
Noncommutative space has been found to be of use in a number of different contexts. In particular, one may use noncommutative spacetime to generate quantised gravity theories. Via an identification between the Moyal $\star$-product on function space and commutators on a Hilbert space, one may use the Seiberg-Witten map to generate corrections to such gravity theories. However, care must be taken with the derivation of commutation relations. We examine conditions for the validity of such an approach, and motivate the correct form for polar noncommutativity in $\mathbb{R}^{2}$. Such an approach lends itself readily to extension to more complicated spacetime parametrisations.
\end{abstract}


\maketitle


\section{Introduction}
The search for a consistent quantum theory of gravity is the predominant focus of much of the physics community. Taking inspiration from the operator formalism in quantum mechanics, one may consider a model of gravity in which the underlying spacetime is quantised:
\begin{equation}
[\hat{x}_{i},\,\hat{x}_{j}]=i\theta_{ij},
\end{equation}
for some antisymmetric matrix of constants $\theta_{ij}$. A theory on this deformed spacetime bears relation to a similar theory on commutative spacetime, where the notion of ordinary functional multiplication is replaced by a deformed product, known as the Moyal $\star$-product. This product forms the basis for the cornerstone of most works on noncommutative spacetime: the Seiberg-Witten map \cite{Seiberg:1999}. While a general gravity theory is not a gauge theory (with the exception of in three dimensions, where the identification between the gravity theory and Chern-Simons theory is well-defined \cite{Achucarro:1986}) and so does not directly satisfy the requirements for the Seiberg-Witten map to be used, one can view Einstein's gravity as, for example, a Poincar\'{e} gauge theory and then consider the noncommutative extension \cite{Chamseddine:2001}.

The above considerations have motivated the search for noncommutative extensions to interesting gravitational theories: in particular, black hole solutions in noncommutative space hold great interest \cite{Nicolini:2009}. A geometry of interest is that of $AdS_{3}$ and the BTZ black hole, whose spacetime is most readily expressed in terms of polar coordinates. Then, to use the Seiberg-Witten map, one must be sure that the correct noncommutative relations are employed in the field theory.

The question of noncommutativity in polar coordinates has been considered previously (for example, see \cite{Hammou:2002} for a description of `fuzzy' $\mathbb{R}^{3}$). We wish to focus on work in \cite{Park:2013} and \cite{Chang:2008}, where a number of commutation relations were postulated for the fundamental commutator, $[\hat{r},\,\hat{\theta}]$, in polar coordinates. These relations were determined and justified only by a dimensional argument, without any recourse to the validity of the association between the Hilbert space and the function space over which the Moyal $\star$-product acts. While in the case of the BTZ black hole, the simplicity of the metric means that the $\star$-product of the triads $\hat{e}_{\mu}^{a}$ reduces to that of standard multiplication and a perceived singularity at $r=0$ is unimportant to the qualitative results, a systematic analysis of coordinate noncommutativity is important if one were to consider more complicated metrics.

\section{Noncommutativity and the Moyal $\star$-product}
We first consider the relationship between a quantum theory, whose quantisation is determined by non-zero commutation relations, and a theory in a function space whose product is defined in a noncommutative sense. This is the required identification for consistent use of the Seiberg-Witten map.

The Moyal-$\star$ product may be written as follows \cite{Bars:2002}. For spatial coordinates $x_{i}$ and corresponding noncommutative parameters $\theta_{ij}$, we define the $\star$-product of two functions valued in this space as
\begin{equation}\label{eq:Moyal}
f\star g(x)=\left.\text{exp}\left(\frac{i}{2}\theta_{ij}\partial_{i}\partial_{j}^{\prime}\right) f(x)g(x^{\prime})\right|_{x^{\prime}=x}.
\end{equation}

We consider, for the purposes of illustration, a noncommutative theory in $\mathbb{R}^{2}$, giving the standard rectangular spatial commutation relation $[x_{1},\,x_{2}]=i\theta_{12}$. For a generic function on this space, we have the associated Fourier transform
\begin{equation*}
\tilde{f}(\alpha_{1},\,\alpha_{2})=\int\ud^{2}x\,e^{i(\alpha_{1}x_{1}+\alpha_{2}x_{2})}f(x_{1},\,x_{2}),
\end{equation*}
We define an operator $\hat{\mathcal{O}}_{f}(\hat{x}_{1},\,\hat{x}_{2})$ on the space of Hilbert functions of the analogous quantum system:
\begin{equation}
\hat{\mathcal{O}}_{f}(\hat{x}_{1},\,\hat{x}_{2})=\frac{1}{(2\pi)^{2}}\int\ud^{2}\alpha \,U(\alpha_{1},\,\alpha_{2})\tilde{f}(\alpha_{1},\,\alpha_{2}),
\end{equation}
where $U(\alpha_{1},\,\alpha_{2})=\exp\left(-i(\alpha_{1}\hat{x}_{1}+\alpha_{2}\hat{x}_{2})\right)$. The product of two such operators may be calculated, using Baker-Campbell-Hausdorff:
\begin{equation*}
U(\alpha_{1},\,\alpha_{2})U(\beta_{1},\,\beta_{2})=U(\alpha_{1}\beta_{1}+\alpha_{2}\beta_{2}) e^{-\frac{i}{2}\theta_{12}(\alpha_{1}\beta_{2}-\alpha_{2}\beta_{1})}.
\end{equation*}
This map defines the correspondence between function multiplication on a noncommutative space and quantum mechanical commutation relations. To see this, we seek an expression for $\hat{\mathcal{O}}_{f}\hat{\mathcal{O}}_{g}$:
\begin{align*}
\hat{\mathcal{O}}_{f}\hat{\mathcal{O}}_{g}=\frac{1}{(2\pi)^{4}}\int\ud^{2}\alpha\,\ud^{2}\beta \,&U(\alpha_{1},\,\alpha_{2})U(\beta_{1},\,\beta_{2})\tilde{f}(\alpha_{1},\,\alpha_{2})\tilde{g}(\beta_{1},\,\beta_{2}) \\
=\frac{1}{(2\pi)^{4}}\int\ud^{2}\alpha\,\ud^{2}\beta\,& U(\alpha_{1}+\beta_{1},\,\alpha_{2}+\beta_{2})e^{-\frac{i}{2}\theta_{12}(\alpha_{1}\beta_{2}-\alpha_{2}\beta_{1})}\\
&\cdot\tilde{f}(\alpha_{1},\,\alpha_{2})\tilde{g}(\beta_{1},\,\beta_{2}).
\end{align*}
Under a suitable change of variables, this becomes
\begin{align*}
\hat{\mathcal{O}}_{f}\hat{\mathcal{O}}_{g}=\frac{1}{(2\pi)^{4}}\int&\ud^{2}\gamma\,\ud^{2}\delta \,U(\gamma_{1},\,\gamma_{2})e^{\frac{i}{2}\theta_{12}(\gamma_{1}\delta_{2}-\gamma_{2}\delta_{1})} \\
&\cdot\tilde{f}\left(\frac{\gamma_{1}}{2}+\delta_{1},\, \frac{\gamma_{2}}{2}+\delta_{2}\right)\tilde{g}\left(\frac{\gamma_{1}}{2}-\delta_{1},\, \frac{\gamma_{2}}{2}-\delta_{2}\right).
\end{align*}
At this point, we note that the Fourier transform of the Moyal $\star$-product \eqref{eq:Moyal} in two spatial dimensions can be put into the form
\begin{align*}
\widetilde{f\star g}(\gamma_{1},\,\gamma_{2})=\int\ud^{2}\delta\,& e^{\frac{i}{2}\theta_{12}(\gamma_{1}\delta_{2}-\gamma_{2}\delta_{1})}\\
&\cdot\tilde{f}\left(\frac{\gamma_{1}}{2}+\delta_{1},\, \frac{\gamma_{2}}{2}+\delta_{2}\right)\tilde{g}\left(\frac{\gamma_{1}}{2}-\delta_{1},\, \frac{\gamma_{2}}{2}-\delta_{2}\right),
\end{align*} 
and so
\begin{align*}
\hat{\mathcal{O}}_{f}\hat{\mathcal{O}}_{g}=&\frac{1}{(2\pi)^{2}}\int\ud^{2}\gamma\,U(\gamma_{1},\,\gamma_{2})\widetilde{f\star g}(\gamma_{1},\,\gamma_{2}) \\
 =&\hat{\mathcal{O}}_{f\star g}.
\end{align*}
Given this correspondence, we may derive spatial commutation relations of functions on $\mathbb{R}^{2}_{\text{NC}}$ as operator relations on a Hilbert space of operators, and vice versa. The key point is the validity of this transformation: if the product $f\star g$ is divergent, then this correspondence has no meaning and the associated quantum theory may not be valid.
While the $\star$-product has obvious utility, the form outlined in \eqref{eq:Moyal} does not seem to admit tractable results. In fact, this is the case: for all but the simplest parametrisations of the space in question, one must undertake an order-by-order expansion of the exponent and derive the results as a series expansion. Such a procedure is not guaranteed to work, but if it is successful then there are two possible outcomes.
\begin{itemize}
\item The series expansion converges or terminates: In the event that the expansion is equivalent to a series expansion of a well-defined function, then the equivalence is valid and the result of $f\star g$ will provide the details of the commutatation relation(s) in the quantum theory. This possibility includes those cases where the series terminates, and hence provides a convergent result. A reparametrisation to coordinates $f,\,g$ is reasonable;
\item The series fails to converge: While the defined functional noncommutativity is valid, the singularities in the function space prevent the identification from being utilised. In this case, a reparamterisation to coordinates $f,\,g$ for use in the quantum commutation relations will result in a singular or ill-defined Hilbert space in the quantum picture.
\end{itemize}
Note that if we apply the above procedure to the standard rectangular coordinates, $f(x,\,y)=x$ and $g(x,\,y)=y$, the $\star$-product terminates rapidly and leaves us with the commutation relation we started with, namely $[x,\,y]=i\theta$. We now consider the state of affairs when the space is parametrised by polar coordinates.

\section{Polar Noncommutativity}
A brief reflection on the nature of polar coordinates suggests a number of problems that could emerge. Firstly, in commutative space, the identification $x=r\cos\phi$, $y=r\sin\phi$ is unambiguous and so is the converse, $r=\sqrt{x^{2}+y^{2}}$, $\phi=\arctan(y/x)$. However, if one anticipates the commutator $[\hat{r},\,\hat{\phi}]\neq0$, then the ordering of the terms in the reparametrisation becomes important, and it is not at all clear what the meaning of this identification is. Indeed, in \cite{Chang:2008}, the commutation relation derived between $\hat{r}$ and $\hat{\phi}$ resists any meaningful interpretation due to its $\hat{r}^{-1}$ dependence. Due to the singularity at the origin, this suggests that the commutator is not valid over the whole of $\mathbb{R}^{2}$. A sensible approach to derive such a relation is to start with the well-understood relation
\begin{equation}
[z,\,\bar{z}]=2i\theta,
\end{equation}
express $z=re^{i\phi}$ and Taylor expand the exponent to isolate $[r,\,\phi]_{\star}$. However, it does not take long to see that there is no clear way to extract the commutator from its surroundings, and it appears that the result asserted previously applies only to first-order. To avoid any such ambiguities, and to formalise the analysis, we begin from commutative space with the Moyal $\star$-product and consider convergence of the results gained. It transpires that the commutator $[\hat{r},\,\hat{\phi}]_{\star}$ has no meaning in the context of the Moyal identification due to its clear divergence.

Beginning with the proffered `simplest' commutation relation in polar coordinates, we select $f(x,\,y)=r\equiv\sqrt{x^{2}+y^{2}}$ and $g(x,\,y)=\phi\equiv\arctan(y/x)$, with underlying rectangular commutation relation $[x,\,y]=i\theta$. Expanding the Moyal $\star$-product in derivatives to $\mathcal{O}(\theta^{40})$, one may deduce the result:
\begin{equation}
[r,\,\phi]_{\star}\equiv r\star\phi-\phi\star r=-\frac{i\theta}{r}\sum_{n=0}^{\infty}\frac{(4n)!}{(2n+1)!}\left(\frac{\theta}{4r^{2}}\right)^{2n}.
\end{equation}
A cursory examination of this result is enough to determine that this is not convergent for any values of $r$ and $\theta$ except for the trivial cases: $\theta=0$ (commutative) or $r\to\infty$ (pseudo-commutative). While the first-order result yields $[r,\,\phi]=-i\theta/r$, in agreement with the assertions in \cite{Chang:2008}, it is clear that this first-order result is not the dominant term in the expansion of the $\star$-product, and one should not rely on such a result for any analysis of noncommutative gravity theories.

If, then, the expected `simplest' polar commutation relation fails in its regularity, how does one proceed with any meaningful evaluation of noncommutative gravitational theories in polar coordinates? The initial comments on polar coordinates provide an answer: we should aim to eliminate some aspect of the underlying ambiguity in the reparametrisation. Instead, considering the commutator $[r^{2},\,\phi]$, we find the Moyal $\star$-product series expansion vanishes identically at orders higher that $\mathcal{O}(\theta)$, and so the commutator is regular and well-defined:
\begin{equation}\label{eq:FUNDAMENTAL}
[r^{2},\,\phi]_{\star}=2i\theta.
\end{equation}
This was the relation eventually used in \cite{Chang:2008} for its convenience, and so despite the faulty starting point the results derived therein maintain their validity. However, it is important to point out that \eqref{eq:FUNDAMENTAL} is the fundamental commutation relation for polar noncommutative $\mathbb{R}^{2}$. Intuitively, such a result makes sense: the coordinate system is poorly defined for $r=0$, and so it is unsurprising that the naive commutation relation resulted in a problematic $r^{-1}$ dependence. Around $r=0$, the more natural way to consider the coordinates $(x,\,y)$ is as equivalent to $(r^{2},\,\phi)$. This gives some heuristic validation of the above. The more general considerations, however, shed some light on the nature of possible reparametrisations in noncommutative gravity theories.

\section{Conclusions}
In this paper, we have examined the validity of reparametrisations in noncommutative field theories, with regards to their application to quantised gravity theories. It is important that care be used when choosing such a reparametrisation, as the implicit equivalence that the Moyal $\star$-product provides between function space and Hilbert space comes with important caveats on the behaviour and convergence of the $\star$-product results. In particular, when considering polar noncommutativity, the correct parameters to consider are seen to be $r^{2}$ and $\phi$ as they provide a convergent (in fact, terminating) series on the $\star$-product side which translates into a regular, well-defined commutation relation on the quantum side.

This may have implications for other attempts to construct noncommutative quantum theories, particularly in the nature of reparametrisations. For example, if one were to choose a set of commutation relations for $\mathbb{R}^{3,\,1}$ and compute the metric for a noncommutative Schwarzchild black hole, then the coordinate singularities evinced in the standard spacetime coordinate description may no longer be trivial to remove. As discussed in \cite{Chang:2008}, the BTZ black hole in noncommutative spacetime appears to demonstrate the presence of additional horizons; this feature may be a generic feature of any such gravity theory. However, the intricacies of employing the Seiberg-Witten map in these contexts must also be reckoned with. Nevertheless, a rigorously defined quantum mechanical correspondence is a necessity for any such analysis, as we have demonstrated.

\section*{Acknowledgements}
The author would like to thank Douglas Smith and Henry Maxfield for helpful feedback and comments, and James Edwards for useful discussion and proofreading. The author is supported by an EPSRC studentship.

\bibliography{referencespolar}

\begin{thebibliography}{8}
\providecommand{\natexlab}[1]{#1}
\providecommand{\url}[1]{\texttt{#1}}
\providecommand{\urlprefix}{URL }
\expandafter\ifx\csname urlstyle\endcsname\relax
  \providecommand{\doi}[1]{doi:\discretionary{}{}{}#1}\else
  \providecommand{\doi}[1]{doi:\discretionary{}{}{}\begingroup
  \urlstyle{rm}\url{#1}\endgroup}\fi
\providecommand{\bibinfo}[2]{#2}

\bibitem[{Seiberg and Witten(1999)}]{Seiberg:1999}
\bibinfo{author}{N.~Seiberg}, \bibinfo{author}{E.~Witten},
  \bibinfo{title}{String theory and noncommutative geometry},
  \bibinfo{journal}{Journal of High Energy Physics}
  \bibinfo{volume}{1999}~(\bibinfo{number}{09}) (\bibinfo{year}{1999})
  \bibinfo{pages}{032}.

\bibitem[{Ach{\'u}carro and Townsend(1986)}]{Achucarro:1986}
\bibinfo{author}{A.~Ach{\'u}carro}, \bibinfo{author}{P.~K. Townsend},
  \bibinfo{title}{A {C}hern-{S}imons action for three-dimensional anti-de
  {S}itter supergravity theories}, \bibinfo{journal}{Physics Letters B}
  \bibinfo{volume}{180}~(\bibinfo{number}{1}) (\bibinfo{year}{1986})
  \bibinfo{pages}{89--92}.

\bibitem[{Chamseddine(2001)}]{Chamseddine:2001}
\bibinfo{author}{A.~H. Chamseddine}, \bibinfo{title}{Deforming {E}instein's
  gravity}, \bibinfo{journal}{Physics Letters B}
  \bibinfo{volume}{504}~(\bibinfo{number}{1}) (\bibinfo{year}{2001})
  \bibinfo{pages}{33--37}.

\bibitem[{Nicolini(2009)}]{Nicolini:2009}
\bibinfo{author}{P.~Nicolini}, \bibinfo{title}{Noncommutative black holes, the
  final appeal to quantum gravity: a review}, \bibinfo{journal}{International
  Journal of Modern Physics A} \bibinfo{volume}{24}~(\bibinfo{number}{07})
  (\bibinfo{year}{2009}) \bibinfo{pages}{1229--1308}.

\bibitem[{Hammou et~al.(2002)Hammou, Lagraa, and Sheikh-Jabbari}]{Hammou:2002}
\bibinfo{author}{A.~Hammou}, \bibinfo{author}{M.~Lagraa},
  \bibinfo{author}{M.~Sheikh-Jabbari}, \bibinfo{title}{Coherent state induced
  star product on R $\lambda$ 3 and the fuzzy sphere},
  \bibinfo{journal}{Physical Review D}
  \bibinfo{volume}{66}~(\bibinfo{number}{2}) (\bibinfo{year}{2002})
  \bibinfo{pages}{025025}.

\bibitem[{Park(2013)}]{Park:2013}
\bibinfo{author}{M.-I. Park}, \bibinfo{title}{The rotating black hole in
  renormalizable quantum gravity: The three-dimensional {H}o{\v{r}}ava gravity
  case}, \bibinfo{journal}{Physics Letters B}
  \bibinfo{volume}{718}~(\bibinfo{number}{3}) (\bibinfo{year}{2013})
  \bibinfo{pages}{1137--1141}.

\bibitem[{Chang-Young et~al.(2009)Chang-Young, Lee, and Lee}]{Chang:2008}
\bibinfo{author}{E.~Chang-Young}, \bibinfo{author}{D.~Lee},
  \bibinfo{author}{Y.~Lee}, \bibinfo{title}{{Noncommutative BTZ Black Hole in
  Polar Coordinates}}, \bibinfo{journal}{Class.Quant.Grav.}
  \bibinfo{volume}{26} (\bibinfo{year}{2009}) \bibinfo{pages}{185001},
  \doi{\bibinfo{doi}{10.1088/0264-9381/26/18/185001}}.

\bibitem[{Bars and Matsuo(2002)}]{Bars:2002}
\bibinfo{author}{I.~Bars}, \bibinfo{author}{Y.~Matsuo},
  \bibinfo{title}{Computing in string field theory using the {M}oyal star
  product}, \bibinfo{journal}{Physical Review D}
  \bibinfo{volume}{66}~(\bibinfo{number}{6}) (\bibinfo{year}{2002})
  \bibinfo{pages}{066003}.

\end{thebibliography}
\bibliographystyle{elsarticle-num-names}

\end{document}